\documentclass[pre,10pt,twocolumn,superscriptaddress,floatfix,showpacs]{revtex4-1}
\usepackage[utf8]{inputenc}  %Input what you want e.g., é, ł, a, ü
\usepackage[T1]{fontenc}     %Output what you want e.g., é, ł, a, ü
\usepackage[british]{babel}  %Do hyphenation according to british english
\usepackage[sc,osf]{mathpazo}
\usepackage[scaled=0.86]{berasans}  % URL font that go well wtih palatino
\usepackage[colorlinks=true, citecolor=blue, urlcolor=blue]{hyperref}  %Hyperlinks (pink, green, blue)
\usepackage{graphicx} % Package to insert exteral figures
\usepackage[babel]{microtype}  %Improves text justification
\usepackage{amsmath,amssymb,amsthm,bm,amsfonts,mathrsfs,bbm} %Usefull math packages
\usepackage{xspace}  %Useful to add space in macros
\usepackage{pgfplots}

\DeclareMathOperator*{\Desc}{Des}

\begin{document}
\title{The presence of quantum correlations result in non-vanishing ergotropic gap}

\author{Amit Mukherjee}
\email{amitisiphys@gmail.com}
\affiliation{Physics and Applied Mathematics Unit, Indian Statistical Institute, 203 B. T. Road, Kolkata 700108, India.}

\author{Arup Roy}
\email{arup145.roy@gmail.com}
\affiliation{Physics and Applied Mathematics Unit, Indian Statistical Institute, 203 B. T. Road, Kolkata 700108, India.}

\author{Some Sankar Bhattacharya}
\email{somesankar@gmail.com}
\affiliation{Physics and Applied Mathematics Unit, Indian Statistical Institute, 203 B. T. Road, Kolkata 700108, India.}

\author{Manik Banik}
\email{manik11ju@gmail.com}
\affiliation{Optics \& Quantum Information Group, The Institute of Mathematical Sciences, C.I.T Campus, Tharamani, Chennai 600 113, India.}

%%%%%%%%%%%%%%%%%%%%%%%%%%%%%%%%%%%%%%%%%%%%%%%%%%%%%%%%%%%%%%%%%%%%%%%%%%%%%%%%%%%%%%%%%%%%%%%%
%%%%%%%%%%%%%%%%%%%%%%%%%%%%%%%%%%%%%%%%%%%%%%%%%%%%%%%%%%%%%%%%%%%%%%%%%%%%%%%%%%%%%%%%%%%%%%%%

\begin{abstract}
The paradigm of extracting work from isolated quantum system through a cyclic Hamiltonian process is a topic of immense research interest. The optimal work extracted under such process is termed as \emph{ergotropy} [\href{http://iopscience.iop.org/0295-5075/67/4/565}{Europhys. Lett., {\bf 67} (4), 565(2004)}]. Here, in a multi-party scenario we consider only a class of such cyclic processes that can be implemented locally, giving rise to the concept of \emph{local ergotropy}. Eventually, presence of quantum correlations result in a non-vanishing thermodynamic quantity called \emph{ergotropic gap}, measured by the difference between the global and local ergotropy. However the converse does not hold in general, i.e. its nonzero value does not necessarily imply presence of quantum correlations. For arbitrary multi-party states we quantify this gap. We also evaluate the difference between maximum global and local extractable work for arbitrary states when the system is no longer isolated but put in contact with a baths of same \emph{local} temperature. 
\end{abstract}
\maketitle 

%%%%%%%%%%%%%%%%%%%%%%%%%%%%%%%%%%%%%%%%%%%%%%%%%%%%%%%%%%%%%%%%%%%%%%%%%%%%%%%%%%%%%%%%%%%%%%%%
%%%%%%%%%%%%%%%%%%%%%%%%%%%%%%%%%%%%%%%%%%%%%%%%%%%%%%%%%%%%%%%%%%%%%%%%%%%%%%%%%%%%%%%%%%%%%%%%

\section{Introduction}
The idea of information is deeply connected with Physics, especially with thermodynamics\cite{Landauer'1961,Bennett'1982,Vedral'2009}. Considerable effort has been devoted to ameliorate this connection in the quantum regime \cite{Lubkin'1987,Vedral'2011}. In consequence, resource-theoretic aspects of quantum thermodynamics have flourished \cite{Spekkens'2013,Oppenheim'2013,Wehner'2014,Gour'2015}. Although the importance of quantum correlations in the context of quantum thermodynamics is not understood in full generality, till date, but substantial amount of progress has been made in this direction in recent past \cite{Fannes'2013,Acin'2013,Skrzypczyk'2014}.

The presence of correlations that have no classical counterpart is one of the striking features of multi-party quantum systems. One much studied way to capture the notion of quantumness present in a correlation is \emph{entanglement} \cite{Horodecki'2009}. However, there exist several tasks where a multi-party quantum state, not being entangled at all, can be more advantageous than classical correlation. In bipartite scenario this quantumness present in a correlation has been quantified by the quantity known as \emph{discord}, found to be a useful resource for various information theoretic tasks viz. extended state merging \cite{Madhok'2011,Cavalcanti'2011}, remote state preparation \cite{Dakic'2012}, although in a restricted sense. This initiated the study of quantumness in a more general framework than entanglement \cite{Vedral'2012}. In this work we investigate whether quantumness in correlations have some implications in quantum thermodynamics. Interestingly, we show that there exists a thermodynamic quantity namely, the \emph{ergotropic gap}, the difference between maximum extractable work under global and local cyclic Hamiltonian process, is non vanishing whenever the multi-party quantum state is not classically correlated.

Extracting work from quantum system is one of the important areas of study in quantum thermodynamics \cite{Lifshitz'1978,Callen'1985}. The question of optimal work extraction from an isolated quantum system under cyclic Hamiltonian process was first studied in the mathematical framework of $C^*$-algebra \cite{Woronowicz'1978} which was later explored in well known Hilbert space formalism \cite{Lenard'1978}. The aim is to transform a quantum system from a higher to a lower internal energy state, extracting the difference in internal energy as work. It has been shown that optimal amount of work is extracted under a cyclic Hamiltonian process whenever the system evolves into a state, called \emph{passive} state \cite{Woronowicz'1978,Lenard'1978,Nieuwenhuizen'2004}, from which no further work can be extracted. The authors of ref.\cite{Nieuwenhuizen'2004} coined the term \emph{ergotropy} for optimal extractable work.

In recent past the topic of extracting work from quantum system has gained renewed interest \cite{Aberg'2013,Horodecki'2013,Popescu'2014,Aberg'2014,Lostaglio'2015,Acin'2014}. In \cite{Acin'2014}, the authors designed a scenario where correlation in multi-party quantum system enables work extraction. Given a non-interacting Hamiltonian of a multi-party system, any cyclic unitary process can be realized by switching on a suitable external interaction field. Here we consider a situation where subsystems are spatially separated and no global external interacting field can be implemented on the total system. Each subsystem can only be acted upon by a local field. We call the optimal extractable work \emph{local ergotropy}. We find that there exists classically correlated states for which ergotropic gap, the difference between global and local ergotropy, can be nonzero. But this does not lead to the conclusion that classical correlations always possess non zero \emph{ergotropic gap} because there exist classically correlated states for which this gap turns out to be zero. Interestingly, we find that whenever the multi-party system is not classically correlated then the optimal amount of extractable work under cyclic local interaction is strictly less than that obtained under global interaction, i.e. presence of quantum correlations always result in non vanishing \emph{ergotropic gap}. Given a non-interacting Hamiltonian and arbitrary initial state of a multi-party system we quantify this gap. We also consider the scenario where the system is no longer isolated but put in contact with a baths of the same \emph{local} temperature and evaluate the difference between maximum global and local extractable work for arbitrary state.

\section{framework for work extraction}
Consider a quantum system, composed of $N$ subsystems, prepared in the state $\varrho_{_{A_1...A_N}}\in\mathcal{D}(\mathcal{H}_1\otimes...\otimes\mathcal{H}_N)$, where $\mathcal{H}_i$ denotes the Hilbert space corresponding to the $i^{th}$ subsystem and $\mathcal{D(\mathcal{X})}$ denotes the set of density operator acting on Hilbert space $\mathcal{X}$. Consider that the local Hamiltonian for the $i^{th}$ party is given by $H_{i}=\sum_{j}e^{j}_i|j_i\rangle\langle j_i|$, where $|j_i\rangle$ denotes the $j^{th}$ energy eigenstate of the $i^{th}$ particle with energy eigenvalue $e^{j}_i$. No interactions are considered among the various subsystems. So, the total Hamiltonian of the composite system takes the form:
\begin{equation}\label{Hamil} 
H=\displaystyle\sum_{i=1}^NH_i\bigotimes_{k\in\bar{i}}
\mathbf{I}_k=\sum\limits_{i=1}^{N}\tilde{H}_{i},
\end{equation} 
where $H_i\bigotimes_{k\in\bar{i}}\mathbf{I}_k=\mathbf{I}_1\otimes...\otimes\mathbf{I}_{i-1}\otimes H_i\otimes\mathbf{I}_{i+1}\otimes...\otimes\mathbf{I}_N$ with $\mathbf{I}_{l}$ denoting identity operator acting on the Hilbert space $\mathcal{H}_l$. 

In the paradigm of work extraction from an isolated system under a cyclic unitary process the protocol is to transform the state from $\varrho_{_{A_1...A_N}}$ to some $\sigma_{_{A_1...A_N}}$ by using some time dependent unitary operation $U(\tau)$ such that $\sigma_{_{A_1...A_n}}$ has less internal energy than $\varrho_{_{A_1...A_N}}$. Note that, since only unitary operations are used, the entropy of the final state is same as the initial. Any such unitary can be generated by applying a time dependent interaction $V(t)$ among the $N$ subsystems, such that $V(t)$ is non-vanishing only when $0 \leq t \leq \tau$. The corresponding evolution can be described by the unitary operator $U(\tau)= \overrightarrow\exp\left(-\iota\int_0^\tau d t\left(H + V(t)\right)\right)$, where $\overrightarrow\exp$ denotes the time-ordered exponential. In this setup the optimally extractable work is therefore:
\begin{eqnarray*} 
W^G_{opt}&=& \max_{U(\tau)}\mbox{Tr}[(\varrho_{_{A_1...A_N}}-U(\tau)\varrho_{_{A_1...A_N}}U^{\dagger}(\tau))H],\\
&=& \mbox{Tr}[\varrho_{_{A_1...A_N}}H]-\min_{U(\tau)}\mbox{Tr}[U(\tau)\varrho_{_{A_1...A_N}}U^{\dagger}(\tau)H],
\end{eqnarray*} 
where optimization is done over all unitaries. It has been shown that this optimization makes the system evolve into a state $\varrho^{passive}_{_{A_1...A_N}}$, called passive state \cite{Woronowicz'1978,Lenard'1978,Nieuwenhuizen'2004}. Thus the optimal amount of extractable work, namely ergotropy \cite{Nieuwenhuizen'2004}, amounts to
\begin{equation}\label{WG}
W^G_{opt}=\mbox{Tr}[\varrho_{_{A_1...A_N}}H]-\mbox{Tr}[\varrho^{passive}_{_{A_1...A_N}}H].
\end{equation} 
Note that among the passive states there is a special one, called thermal or Gibbs state. Given many copies of the system it may be possible that work can be extracted even from passive states. But no such work extraction is possible in case of thermal states, so it is called \emph{completely} passive state \cite{Nieuwenhuizen'2004,Brunner'2015}. 

Now consider a situation where each sub-system of the joint system $\varrho_{_{A_1...A_N}}$ is spatially separated and implementation of any global interaction field is not allowed. Each party can only apply time dependent local field on their respective subsystem. Hence the interaction on the composite system looks
\begin{equation}\label{int}
V(t)=\displaystyle\sum_{i=1}^NV_i(t)
\bigotimes_{k\in\bar{i}}\mathbf{I}_k=
\sum\limits_{i=1}^{N}\tilde{V}_{i}(t).
\end{equation}  
The class of unitaries generated from such interactions look
\begin{eqnarray}\nonumber
U(\tau)&=&\overrightarrow\exp\left(  -\iota\int_0^\tau d t \sum\limits_{i=1}^{N}\left(\tilde{H}_i + \tilde{V}_i(t)\right)\right)\\\nonumber
&=&\prod\limits_{i=1}^{N}\overrightarrow\exp\left(  -\iota\int_0^\tau d t~~\{H_i + V_i(t)\}\bigotimes_{k\in\bar{i}}\mathbf{I}_k\right)\\\nonumber
&=&\prod\limits_{i=1}^{N}\left\lbrace \overrightarrow\exp\left(  -\iota\int_0^\tau d t~~\{H_i + V_i(t)\}\right)\bigotimes_{k\in\bar{i}}\mathbf{I}_k\right\rbrace\\\nonumber
&=& \bigotimes_{i=1}^{N}\overrightarrow\exp\left(-\iota\int_0^\tau d t\left(H_i + V_i(t)\right)\right)\\\nonumber
&=& \bigotimes_{i=1}^{N}U_i(\tau).
\end{eqnarray}
where $U_i(\tau)$ is the unitary on $i^{th}$ particle. Here we ignore global constant factor which is not relevant. Let us denote this class of unitaries as
\begin{equation}\label{localu}
\mathcal{LU}:=\left\lbrace U(\tau)~|~U(\tau)=\bigotimes_{i=1}^{N}U_i(\tau)\right\rbrace ,
\end{equation} 
The optimal work that can be extracted under such local interactions is thus 

\begin{eqnarray}\label{WL}  
W_{opt}^L:&=&\max\limits_{U\in\mathcal{LU}}\mbox{Tr}[(\varrho_{_{A_1...A_n}}-U\varrho_{_{A_1...A_n}}U^\dagger)H]\nonumber\\
&=& \mbox{Tr}[\varrho_{_{A_1...A_n}}H]-\min\limits_{U\in\mathcal{LU}}\mbox{Tr}[U\varrho_{_{A_1...A_n}}U^\dagger H]. 
\end{eqnarray}
In this scenario since work is extracted under applying local unitaries, we call the optimal extractable work \emph{local ergotropy} of the state $\varrho$ given the Hamiltonian $H$. In the above notation superscript $L$ is introduced to distinguish this quantity from the one defined in Eq.(\ref{WG}), where the superscript $G$ has been used to denote that global unitaries are allowed. 

At this point we define a quantity which is the difference of the global and local ergotropy, called ergotropic gap(EG)
\begin{equation}\label{eq6}
W_{EG}:=W_{opt}^G-W_{opt}^L.
\end{equation}
Replacing $W_{opt}^G$ and $W_{opt}^L$ from Eq.(\ref{WG}) and Eq.(\ref{WL}) respectively, we have  
\begin{equation} 
W_{EG}=\min\limits_{U\in\mathcal{LU}}\mbox{Tr}[U\varrho_{_{A_1...A_n}}U^\dagger H]- Tr[\rho ^{passive}_{_{A_1...A_n}}H].
\end{equation}
It is easy to see that $W_{EG}$ can not be negative. The reason behind this non negativity can be explained in the following way, local operations are restricted to extract energy from subsystems only where as global unitary can have the power to extract energy from subsystems as well as from correlations. In the following we study this quantity in presence of some amount of correlations between the subsystems of multiparty systems.

\section{Correlation and Ergotropic Gap}
In physics the study of correlation is quite important as it is the most significant feature to characterize multi-particle systems. However its characterization and quantification becomes notoriously difficult when one shifts from classical realm to quantum realm. The core interest of quantum information theory is to study these correlations which are also important from a foundational perspective. Depending on different situations correlations can be characterized in different ways, \emph{eg.} \emph{nonlocal} \cite{Brunner'2014}, \emph{steerable} \cite{Wiseman'2007}, \emph{entanglement} \cite{Horodecki'2009}, \emph{quantum correlation (discord)} \cite{Vedral'2012} \emph{etc} that find number of practical applications \cite{application1,application2,application3,
application4,application5,application6,application7,
application8,application9,application10,application11,
application12, Madhok'2011,Cavalcanti'2011,Dakic'2012}. It also plays important role in quantum thermodynamics \cite{Acin'2014}. Here we are interested in the role of correlation in \emph{ergotropy}.   

An $N$-particle quantum state, with $d$-levels for each particle ($\varrho_{_{A_1...A_n}}$) has $d^N$ eigenvalues (possibly degenerate) forming a normalized probability vector $(\lambda)^{d^N}_1$, represented in a row. Rearrange the eigenvalues and form a vector $\vec{\lambda}=(\lambda_{\alpha})_{\alpha=1}^{d^N}$ where $\lambda_{\alpha}\ge\lambda_{\alpha+1}~\forall~\alpha$. Denote the $d^N$ energy eigenstates (there may be degeneracy) of the Hamiltonian of Eq.(\ref{Hamil}) as $\{|\xi_{\alpha}\rangle\}_{\alpha=1}^{d^N}$ with energy eigenvalues $\xi_{\alpha}\le\xi_{\alpha+1}~\forall~\alpha$. In this notation the passive state $\varrho_{_{A_1...A_n}}^{passive}$ reads as \cite{Woronowicz'1978,Lenard'1978,Nieuwenhuizen'2004},
\begin{equation}
\varrho_{_{A_1...A_n}}^{passive}:=\sum_{\alpha}\lambda_{\alpha}|\xi_{\alpha}\rangle\langle\xi_{\alpha}|.
\end{equation}
$\varrho_{_{A_1...A_n}}^{passive}$ commutes with the Hamiltonian which is diagonalizable in the orthonormal product basis (ONPB) $\{\bigotimes_i|j_i\rangle\}_{j,i}$ where $\{|j_i\rangle\}_j$ forms a orthonormal basis (ONB) (energy eigenbasis) of the $i^{th}$ party Hilbert space $\mathcal{H}_i$. If the multi-particle system is in pure product state then EG is always zero. Consider such an arbitrary state, 
\begin{equation}
\rho_{_{A_1...A_n}}^{product}=|\psi\rangle_{A_{1}} \otimes |\psi\rangle_{A_{2}} ... \otimes |\psi\rangle_{A_{n}},
\end{equation}
where $|\psi\rangle_{A_{i}}\in \mathbb{C}^{d_i},~\forall~i$. Let the ground energy state of the $i^{th}$ particle is $|0\rangle_{A_{i}}$. Applying local unitaries the state of the each subsystem can be transformed from $|\psi\rangle_{A_{1}}$ to $|0\rangle_{A_{i}}$ resulting the global state into its passive form $\varrho_{_{A_1...A_n}}^{passive}=\bigotimes_i |0\rangle_{A_{i}}$. It readily follows that the EG for pure product states are vanishing. Now we ask whether EG of correlated states are vanishing or not. We first start with CC states.

 An $N$-particle state is called classically correlated (CC) if it can be written as \cite{Groisman'2007}
\begin{equation}
\varrho_{_{CC}}=\displaystyle\sum\limits_{\{\beta_i\}\in ONB[\mathcal{H}_i]} p_{\beta_1...\beta_N}\bigotimes_{i=1}^N |\beta_i\rangle\langle \beta_i|,
\end{equation}
where $\{|\beta_i\rangle\}_{\beta}$ is an ONB for the $i^{th}$ particle Hilbert space $\mathcal{H}_i$, and  $(p_{\beta_1...\beta_N})_1^{d^N}$ is a probability vector. Clearly the state $\varrho_{_{CC}}$ is  diagonalized in the ONPB $\{\bigotimes_i|\beta_i\rangle\}_{\beta,i}$. Consider such a two-qubit CC state of the following form, 
\begin{equation}
\varrho_{_{A_1A_2}}=\lambda_1|0\rangle_{A_1}\langle 0|\otimes |0\rangle_{A_2}\langle 0|+\lambda_2|1\rangle_{A_1}\langle 1|\otimes |1\rangle_{A_2}\langle 1|,
\end{equation}
where $0<\lambda_2<\lambda_1<1;~\lambda_1+\lambda_2=1$, and $|0\rangle_{A_i} (|1\rangle_{A_i})$ represents the ground (excited) energy eigenstate of the $i^{th}$ particle Hamiltonian $H_i=e^0|0\rangle_{A_i}\langle 0|+e^1|1\rangle_{A_i}\langle 1|$, with $e^0,e^1$ denoting ground and excited energy eigenvalues, respectively. 
Here the Hamiltonian for the composite system is $H=H_1\otimes\mathbf{I}_2+\mathbf{I}_1\otimes H_2$. The corresponding passive state reads,
\begin{equation}
\varrho_{_{A_1A_2}}^{passive}=\lambda_1|0\rangle_{A_1}\langle 0|\otimes |0\rangle_{A_2}\langle 0|+\lambda_2|0\rangle_{A_1}\langle 0|\otimes |1\rangle_{A_2}\langle 1|.
\end{equation}
For evolving the state $\varrho_{_{A_1A_2}}$ into $\varrho_{_{A_1A_2}}^{passive}$ one needs to apply the  following unitary,
\begin{eqnarray}
|0\rangle_{A_1}\otimes|0\rangle_{A_2}\longmapsto|0\rangle_{A_1}\otimes|0\rangle_{A_2},\nonumber\\
|1\rangle_{A_1}\otimes|1\rangle_{A_2}\longmapsto|0\rangle_{A_1}\otimes|1\rangle_{A_2},
\end{eqnarray}
which is the inverse of the C-Not operation and hence an entangling unitary and can not be realized by unitaries of the form $U_{A_1}\bigotimes U_{A_2}$.

Naturally the question arises whether all CC states possess a non vanishing EG. However, the following example shows that this is not the case in general. Consider the class of CC states of the form, 
\begin{eqnarray}
\rho_{AB}&=& p_0 |0\rangle_{A_{1}} \langle 0|\otimes |0\rangle_{A_{2}} \langle 0|+p_1 |0\rangle_{A_{1}} \langle 0|\otimes |1\rangle_{A_{2}} \langle 1|\nonumber\\ 
&&~~~~~~~~~~~~~+p_2 |1\rangle_{A_{1}} \langle 1|\otimes |0\rangle_{A_{2}} \langle 0|\nonumber\\ 
&&~~~~~~~~~~~~~~~~~~~~+p_3 |1\rangle_{A_{1}} \langle 1|\otimes |1\rangle_{A_{2}} \langle 1|,
\end{eqnarray}
 where $p_0<p_1<p_2<p_3$, $0\le p_k\le 1;~\forall~ k$ and $\sum_{k=0}^3p_k=1$. Corresponding passive state looks,
\begin{eqnarray}
\rho^{passive}_{AB}&=& p_3 |0\rangle_{A_{1}} \langle 0|\otimes |0\rangle_{A_{2}} \langle 0|+p_2 |0\rangle_{A_{1}} \langle 0|\otimes |1\rangle_{A_{2}} \langle 1|\nonumber\\
&&~~~~~~~~~~~~~+p_1 |1\rangle_{A_{1}} \langle 1|\otimes |0\rangle_{A_{2}} \langle 0|\nonumber\\ 
&&~~~~~~~~~~~~~~~~~~~~+p_0 |1\rangle_{A_{1}} \langle 1|\otimes |1\rangle_{A_{2}} \langle 1|.
 \end{eqnarray}
Given such CC states, they can be transformed into passive form by applying $\sigma_x$ operation on each site locally which thus implies vanishing EG. Thus for CC states EG can be zero as well as non-zero.  Here we ask the question whether there exists any correlation possessing non zero EG always. In the following proposition we answer this question.

{\bf Proposition 1:} Ergotropic gap is always non vanishing in presence of quantum correlations.

{\bf Proof:} A quantum state is said to contain quantumness in the correlation if it is not CC, i.e., there is no ONPB that diagonalized the state and such states are called \emph{quantum correlated}. In the bipartite case quantumness is quantified by a quantity called \emph{discord} \cite{Vedral'2012}, which drew a lot of research interest recently.  

It is clear that a quantum correlated state must contain entangled state(s) in its spectrum. However the passive state corresponding to such a state is diagonal in product basis (ONPB of the Hamiltonian). The fact that it is impossible to arrive at some product basis starting from a basis containing entangled state(s) by implementing only local unitaries, implies that $W^{G}_{opt}>W^L_{opt}$ i.e. not all the ergotropy of the system is locally accessible for quantum correlated states.   $~~~~~~~~~~~~~~~~~~~~~~~~~~~~~~~~~~~~~~~~~\blacksquare$

However converse of the above proposition does not hold, i.e., non zero ergotropic gap does not imply presence of quantum correlation as shown in the previous example.
\section{Ergotropic Gap for arbitrary states}
Given the Hamiltonian $H$ of the form of Eq.(\ref{Hamil}) and an arbitrary state it is possible to quantify $W_{EG}$ in terms of the parameters of the Hamiltonian and the state. First we consider two two-level systems and discuss few special sub-classes of states of this system and then we consider multi-party-multi-level systems.  

\subsection{Two-particle-two-level system}
\begin{figure}[t!]
	\centering
	\includegraphics[height=4.5cm,width=5cm]{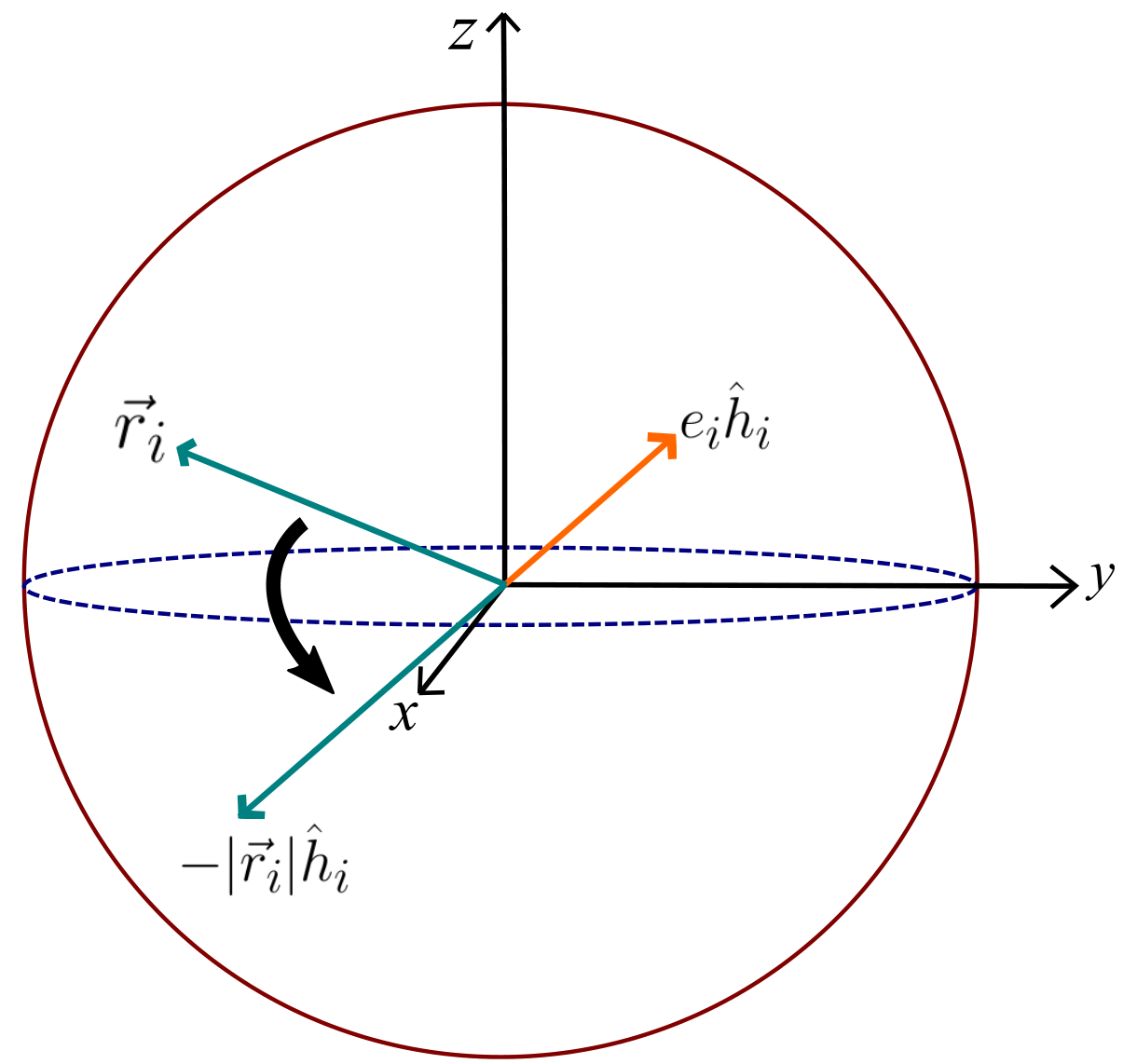}
	\caption{(Color on-line) Bloch sphere for qubit system. $e_i\hat{h}_i$ represents the Bloch vector of the Hamiltonian of Eq.(\ref{hamil22}). For optimizing the second term in Eq.(\ref{WL}) each party apply local unitary $U_i$ that rotates the reduced Bloch vector $\vec{r}_i$ along $-\hat{h}_i$.}\label{fig}
\end{figure}

Consider an arbitrary $2$- particle $2$-level system with Hamiltonian 
\begin{eqnarray}\label{hamil22}
H_i&=& e^0_i|0_i\rangle\langle 0_i|+e^1_i|1_i\rangle\langle 1_i|\nonumber\\
&=&\frac{1}{2}(e^+_i\mathbf{I}+e^-_i\hat{h}_i.\vec{\sigma}),~~i\in{1,2},
\end{eqnarray}
where $e^\pm_i=e^1_i\pm e^0_i$, and $\hat{h}_i$ is a vector in Bloch sphere (see Fig.\ref{fig}) and $\vec{\sigma}\equiv(\sigma_x,\sigma_y,\sigma_z)$, with $\sigma_x,\sigma_y,\sigma_z$ denoting the Pauli matrices. The total Hamiltonian of the composite system is thus
\begin{equation}\label{Hamil2}
H=H_1\otimes\mathbf{I}+\mathbf{I}\otimes H_2.
\end{equation}
An arbitrary two-qubit state can be expressed as the following canonical form \cite{Horodecki'1995}:
\begin{equation}\label{state22} \nonumber
\varrho_{_{A_1A_2}}=\frac{1}{4}\left[ \mathbf{I}\otimes \mathbf{I}+\vec {r_1}.\vec\sigma\otimes\mathbf{I}+\mathbf{I}\otimes\vec{r}_2.\vec\sigma+ \sum\limits_{m,n} t_{mn}\sigma_{m}\otimes\sigma_{n}\right],
\end{equation}
where reduced state of the $i^{th}$ party is 
\begin{equation*}
\varrho_{_{A_i}}=\mbox{Tr}_{A_{\bar{i}}}\left( \varrho_{_{A_1A_2}}\right) =\frac{1}{2}\left[ \mathbf{I}+\vec {r_i}.\vec\sigma\right], 
\end{equation*}
$\vec{r}_i$ being the vectors in $\mathbb{R}^3$ with $|\vec{r}_i|\le 1$.

Let spectral values of $\varrho_{_{A_1A_2}}$ are $\lambda_{00},\lambda_{01},\lambda_{10},\lambda_{11}$, where $\lambda_{00}\ge\lambda_{01}\ge\lambda_{10}\ge\lambda_{11}$. Given the Hamiltonian of Eq.(\ref{Hamil2}) the passive state reads as
\begin{eqnarray}\nonumber
\varrho_{_{A_1A_2}}^{passive}&=&\sum_{x,y}\lambda_{xy}|x_1\rangle\langle x_1|\otimes |y_2\rangle\langle y_2|,
\end{eqnarray} 
with $x,y\in\{0,1\}$. We have 
\begin{equation}\nonumber
\mbox{Tr}\left( \varrho_{_{A_1A_2}}^{passive}H\right) =\sum_{x,y=0}^{1}\lambda_{xy}e_{xy},
\end{equation} 
where $e_{xy}=e^x_1+e^y_2$. Hence we have 
\begin{equation}\nonumber
W^G_{opt}=\mbox{Tr}\left( \varrho_{_{A_1A_2}}H\right) -\left[ \sum\limits_{x,y=0}^{1}\lambda_{xy}e_{xy}\right] ,
\end{equation}
and 
\begin{equation}\nonumber
W^L_{opt}=\mbox{Tr}\left( \varrho_{_{A_1A_2}}H\right) -\min\limits_{U_1\bigotimes U_2}\mbox{Tr}\left[ U\varrho_{_{A_1A_2}}U^\dagger H\right].
\end{equation}
Now observe that,
\begin{eqnarray}\nonumber
\min\limits_{U_1\bigotimes U_2}\mbox{Tr}\left[ U_1\otimes U_2\varrho_{_{A_1A_2}}U_1^\dagger\otimes U_2^\dagger H\right]\nonumber\\
=\min\limits_{U_1\bigotimes U_2}\mbox{Tr}\left[ \varrho_{_{A_1A_2}}U_1^\dagger\otimes U_2^\dagger HU_1\otimes U_2\right] \nonumber\\
=\min\limits_{U_1\otimes U_2}\mbox{Tr}\left[ \varrho_{_{A_1A_2}}U_1^\dagger\otimes U_2^\dagger (H_1\otimes\mathbf{I}+\mathbf{I}\otimes H_2)U_1\otimes U_2\right] \nonumber\\
=\displaystyle\sum\limits_{i=1}^{2}\min\limits_{U_{i}}\mbox{Tr}\left[ \varrho_{_{A_i}}U^\dagger_i H_iU_i\right] \nonumber\\
=\displaystyle\sum\limits_{i=1}^{2}\min\limits_{U_{i}}\mbox{Tr}\left[ U_i\varrho_{_{A_i}}U^\dagger_i H_i\right] .\nonumber
\end{eqnarray}
Therefore we need to independently minimize $\mbox{Tr}[U_i\varrho_{_{A_i}}U^\dagger_i H_i]$ over all $U_i$. To obtain minimum expectation value of the Hamiltonian $H_i$ each party apply the local unitary that rotate the local state-vectors $\vec{r_i}$ along $-\hat{h_i}$ (see Fig.(\ref{fig})). which actually transform the states $\varrho_{_{A_i}}=\frac{1}{2}(\mathbf{I}+\vec{r}_i.\vec{\sigma})$ to the state $\frac{1}{2}(\mathbf{I}-|\vec{r}_i|\hat{h}_i.\vec{\sigma})$. Thus we have,
\begin{eqnarray} 
\min\limits_{U_{i}}\mbox{Tr}\left[ \varrho_{_{A_i}}U^\dagger_i H_iU_i\right]~~~~~~~~~~~~~~~~~~~~~~~~~~~~~~\nonumber\\ 
=\mbox{Tr}\left[ \frac{1}{2}\left( \mathbf{I}- |\vec{r_i}|\hat{h_i}.\vec{\sigma}\right) \frac{1}{2}\left(e^+_i\mathbf{I}+e^-_i\hat{h}_i.\vec{\sigma}\right) \right] \nonumber\\
=\frac{1}{2}\left( e^+_i-e^-_i |\vec{r_i}|\right).\nonumber
\end{eqnarray} 
Hence the local extractable work will be,
\begin{equation}\nonumber
W^L_{opt}=\mbox{Tr}\left( \varrho_{_{A_1A_2}}H\right) -\frac{1}{2}\sum_{i=1}^2\left[e^+_i-e^-_i |\vec{r_i}|\right],
\end{equation}
which further implies,
\begin{equation}
W_{EG}=\frac{1}{2}\sum_{i=1}^2\left[e^+_i-e^-_i |\vec{r_i}|\right]-\left[ \sum\limits_{x,y=0}^{1}\lambda_{xy}e_{xy}\right] .
\end{equation}

For any pure product state, $|\vec{r_1}|=|\vec{r_2}|=1$ and the passive state is $|00\rangle$, i.e., $\lambda_{00}=1$ which, from the above expression, immediately implies $W_{EG}=0$, which is compatible our previous observation that for any pure state ergotropic gap is zero.

\emph{Non zero EG for mixed product state}: Consider a mixed product state of the particular form: $\rho_{AB}=\rho_A\otimes\rho_B=\mbox{diag}\{\alpha,1-\alpha\}\otimes\mbox{diag}\{\beta,1-\beta\}=\mbox{diag}\{\alpha\beta,\alpha(1-\beta),(1-\alpha)\beta,(1-\alpha)(1-\beta)\}$ having same Hamiltonian as in Eq.(\ref{hamil22}). Consider the case $\beta<\alpha<\frac{1}{2}$. The states $\rho_i$ can also be written as $\frac{1}{2}\left[ \mathbf{I}+\vec {r_i}.\vec\sigma\right]$ with $|\vec{r}_1|=1-2\alpha$ and $|\vec{r}_2|=1-2\beta$.
The passive state can then be written as $\lambda_{00}=(1-\alpha)(1-\beta),~\lambda_{01}=(1-\beta)\alpha,
~\lambda_{10}=\beta(1-\alpha),~\lambda_{11}=\alpha\beta$.
To transform the state one need to apply the unitary that has the action $|00\rangle\rightarrow|11\rangle$, $|01\rangle\rightarrow|01\rangle$, $|10\rangle\rightarrow|10\rangle$, and $|11\rangle\rightarrow|00\rangle$. Clearly this is an entangling unitary and can not be realized locally. In this case the EG turns out to be,
\begin{equation}
W_{EG}=(\alpha-\beta)(e_2^0-e_1^0)+(\alpha+\beta)e_1^1+(\beta-\alpha)e_2^1.
\end{equation}
In the following we consider few special classes of correlated states of two-qubit system.

\emph{(a) Mixture of Bell states}: The general form of this class is given by
\begin{equation*}
\varrho_{_{Bell}}=\sum_{i=1}^{4}p_i|\mathcal{B}_i\rangle\langle\mathcal{B}_i|,
\end{equation*}
where $\{|\mathcal{B}_i\rangle\}_{i=1}^4$ are four Bell states (one singlet and three triplets). As one can see these states are already diagonal in the Bell basis with spectral values $\{p_i\}_{i=1}^4$. Suitable global unitary can be considered such that the populations $\{p_i\}_{i=1}^4$ can be arranged in descending order as following: $\{p_{max},p'_{max},p''_{max},p_{min}\}$, where $p_{max}$ is the maximum of $\{p_i\}^4_{i=1}$, $p'_{max}$ being the second maximum and so on such that $p_{max}\ge p'_{max}\ge p''_{max}\ge p_{min}$.

 Since the marginal states are completely mixed, we have,
\begin{eqnarray}
W_{EG}(\varrho_{_{Bell}})=e_1^0 (\dfrac{1}{2}-p_{max}-p'_{max})+e_2^0(\dfrac{1}{2}-p_{max}-p''_{max})\nonumber\\
+e_1^1(\dfrac{1}{2}-p''_{max}-p_{min})+e_2^1(\dfrac{1}{2}-p'_{max}-p_{min}).\nonumber
\end{eqnarray}
For the case $e_1^0=e_2^0=0$ and $e_1^1=e_2^1=1$ it takes the simpler form $W_{EG}(\varrho_{_{Bell}})=p_{max}-p_{min}$.

\emph{(b) Werner class of states}: Generic form of this class is given by 
\begin{equation*}
\varrho_{_{W}}=p|\psi^-\rangle\langle\psi^-|+(1-p)\frac{\mathbf{I}}{2}\otimes\frac{\mathbf{I}}{2},
\end{equation*}
The spectral values are $(\frac{1+3p}{4},\frac{1-p}{4},\frac{1-p}{4},\frac{1-p}{4})$. For this entire class of states the completely mixed marginals imply,
\begin{equation}
W_{EG}(\varrho_{_{W}})=\dfrac{1}{2}p\{(e_1^1-e_1^0)+(e_2^1-e_2^0)\},
\end{equation}
which for the case $e_1^0=e_2^0=0$ and $e_1^1=e_2^1=1$, takes the value $W_{EG}(\varrho_{_{W}})=p$. It is known that Werner class of states contain quantumness in correlation for all values of $p$, except $p=0$, which implies non vanishing ergotropic gap for all values of $p$, except $p=0$.

\subsection{Multi-particle-multi-level systems}
Here we generalize the calculation of two-qubit state for arbitrary states of multi-party systems. Consider $N-$particle state $\varrho_{_{A_1...A_N}}$ and the Hamiltonian $H$ which is of the form of Eq.(\ref{Hamil}). Since there is no interaction term in the Hamiltonian the expression of local ergotropy as in Eq.(\ref{WL}) turns out to be 
\begin{eqnarray}
W_{opt}^L&=&\max\limits_{U\in\mathcal{LU}}\mbox{Tr}\left[ \left( \varrho_{_{A_1...A_n}}-U\varrho_{_{A_1...A_n}}U^\dagger\right) H\right] \nonumber\\
&=&\max\limits_{U\in\mathcal{LU}}\mbox{Tr}\left[ \left( \varrho_{_{A_1...A_n}}-U\varrho_{_{A_1...A_n}}U^\dagger\right) \sum_{i=1}^NH_i\bigotimes_{k\in\bar{i}}\mathbf{I}_k\right] \nonumber\\
&=&\mbox{Tr}\left[ \varrho_{_{A_1...A_n}}\sum_{i=1}^NH_i
\bigotimes_{k\in\bar{i}}\mathbf{I}_k\right]\nonumber\\
&&~~~~~~-\min_{U\in\mathcal{LU}}\mbox{Tr}\left[\left( U \varrho_{_{A_1...A_n}}U^\dagger\right) \sum_{i=1}^NH_i\bigotimes_{k\in\bar{i}}\mathbf{I}_k\right] \nonumber\\
&=&\sum_{i=1}^N\mbox{Tr}\left[ \varrho_{_{A_i}}H_i\right]\nonumber\\
&&-\min_{U\in\mathcal{LU}}
\mbox{Tr}\left[\varrho_{_{A_1...A_n}} U^\dagger\left( \sum_{i=1}^NH_i\bigotimes_{k\in\bar{i}}\mathbf{I}_k\right)U \right],\label{eq12}
\end{eqnarray}
where $\varrho_{_{A_i}}=\mbox{Tr}_{\bar{i}}\left( \varrho_{_{A_1...A_n}}\right) $ is the normalized reduced state of the $i^{th}$ sub-system, here $\mbox{Tr}_{\bar{i}}$ denotes the partial trace over all parties except $i$. The second term on the right hand side of Eq.(\ref{eq12}) can be written as 
\begin{eqnarray}
\min_{U\in\mathcal{LU}}
\mbox{Tr}\left[\varrho_{_{A_1...A_n}} U^\dagger\left( \sum_{i=1}^NH_i\bigotimes_{k\in\bar{i}}\mathbf{I}_k\right)U \right]\nonumber\\
=\min_{\bigotimes_{i=1}^NU_i}\mbox{Tr}\left[\varrho_{_{A_1...A_n}} \bigotimes_{i=1}^NU_i^\dagger\left( \sum_{i=1}^NH_i\bigotimes_{k\in\bar{i}}\mathbf{I}_k\right)\bigotimes_{i=1}^NU_i\right] \nonumber\\
=\min_{\bigotimes_{i=1}^NU_i}\mbox{Tr}\left[\varrho_{_{A_1...A_n}} \left( \sum_{i=1}^NU_i^\dagger H_iU_i\bigotimes_{k\in\bar{i}}\mathbf{I}_k\right)\right] \nonumber\\
=\min_{U_i}\sum_{i=1}^N\mbox{Tr}\left[\varrho_{_{A_i}} \left(U_i^\dagger H_iU_i\right)\right] \nonumber\\
=\sum_{i=1}^N\min_{U_i}\mbox{Tr}\left[\varrho_{_{A_i}} \left(U_i^\dagger H_iU_i\right)\right].\label{eq13}
\end{eqnarray} 
Putting the expression of Eq.(\ref{eq13}) into Eq.(\ref{eq12}) we get
\begin{equation}
W^L_{opt}=\sum_{i=1}^{N}\left[ \mbox{Tr}\left( \varrho_{_{A_i}}H_i\right) -\min\limits_{U_{i}}\mbox{Tr}\left( U_i\varrho_{_{A_i}}U^\dagger_i H_i\right) \right].
\end{equation}
From the above expression it is clear that local ergotropy is the sum of optimal work extracted by each party individually by applying local unitary. Obviously, to extract optimal work each party apply suitable unitaries that transform their reduced density matrix $\varrho_{_{A_i}}$ to the corresponding local passive state $\varrho^{passive}_{_{A_i}}$. Therefore, we have  
\begin{eqnarray}
W^L_{opt}=\sum_{i=1}^{N}\left[\mbox{Tr}\left( \varrho_{_{A_i}}H_i\right) -\mbox{Tr}\left( \varrho^{passive}_{_{A_i}}H_i\right)\right].\label{eq15}
\end{eqnarray}
Substituting Eq.(\ref{WG}) and Eq.(\ref{eq15}) in Eq.(\ref{eq6}) we have,
\begin{equation}\label{EGG}
W_{EG}=\sum_{i=1}^{N}\mbox{Tr}\left[ \varrho^{passive}_{_{A_i}}H_i\right] -\mbox{Tr}\left[ \varrho^{passive}_{_{A_1...A_N}}H\right].
\end{equation} 
Thus for any state $\varrho_{_{A_1...A_N}}$ the ergotropic gap is quantified by the difference of the internal energy of the global passive state from the sum of internal energies of passive state of each parties' reduced state. 

Local ergotropy is the sum of the optimal works extracted by each party, locally. As discussed earlier, switching on the suitable time dependent local interacting field $V_i(t)$, each of the party can transform their initial reduced states to local passive state. However as no global interaction is allowed, the global state $\varrho_{_{A_1...A_N}}$ does not in general evolve into its passive form $\varrho^{passive}_{_{A_1...A_N}}$, rather the global state $\varrho_{_{A_1...A_N}}$ evolves into a state $\eta_{_{A_1...A_N}}$ where $\mbox{Tr}_{\bar{i}}\left( \eta_{_{A_1...A_N}}\right) =\varrho^{passive}_{_{A_i}}~\forall~i$. 
At this point one can further ask the question: whether the state $\eta_{_{A_1...A_N}}$ can be considered as a resource for work extraction under the constraint that no global interaction field among different parties can be applied. Interestingly, given many copies of the state $\eta_{_{A_1...A_N}}$, the answer is yes. This is because the composition of many copies of passive systems may not remain passive and exhibit a form of activation \cite{Woronowicz'1978,Lenard'1978}. It has been shown that the only completely passive state is the Gibbs state (thermal state) from where no further work can be extracted even with many (unbounded) copies \cite{Woronowicz'1978}. So using the activation process work can be extracted until each local passive state transforms into completely passive state, i.e., thermal state. In the following we focus on global and local work extraction from correlated quantum system when it is no longer isolated rather put in contact with thermal bath.
    
\section{Extracting work in presence of thermal bath}
Consider the scenario where the system is no longer isolated but each of the subsystems is put in contact with  baths at the same local temperature $\beta^{-1}$. In this scenario one can again be interested in the maximal work that can be extracted via global unitaries acting jointly on the system and the bath and also the amount of maximal extracted work via local unitaries acting jointly on the subsystems and the local bath. In such scenario, it is well known that the extractable work is upper bounded by the difference between initial and final (thermal) free energies \cite{Spekkens'2013,Horodecki'2013,Popescu'2014}, i.e.,
\begin{equation*}
W^{(cb)}_{opt}=F(\sigma)-F(\sigma^{thermal}),
\end{equation*}
where $F(\sigma)=\mbox{Tr}(H\sigma)-\beta^{-1}S(\sigma)$ and $\sigma^{thermal}=\frac{\exp(-\beta H)}{\mathcal{Z}}$ with $\mathcal{Z}=\mbox{Tr}(\exp(-\beta H))$ being the partition function. The super-index `(cb)' is used to denote that work is extracted in contact of bath.Thus we have
\begin{eqnarray*}
W^{G(cb)}_{opt}=F(\varrho_{_{A_1...A_n}})-F(\varrho_{_{A_1...A_n}}^{thermal}),\\
W^{L(cb)}_{opt}=\sum_{i=1}^n\left[ F(\varrho_{_{A_i}})-F(\varrho_{_{A_i}}^{thermal})\right]. 
\end{eqnarray*}
In such a scenario the difference between global and local extractable work is therefore,
\begin{eqnarray}
\Delta W^{(cb)}=W^{G(cb)}_{opt}-W^{L(cb)}_{opt}. 
\end{eqnarray}
Since we have $H=\sum_{i=1}^nH_i$, which further gives $\varrho_{_{A_1...A_n}}^{thermal}=\bigotimes_i\varrho_{_{A_i}}^{thermal}$, therefore we have 
\begin{eqnarray}
\Delta W^{(cb)}=\beta^{-1}\left[ \sum_{i=1}^nS(\varrho_{_{A_i}})-S(\varrho_{_{A_1...A_n}})\right]. 
\end{eqnarray}
For two-particle system the above expression reduces to the well known \emph{quantum mutual information} \cite{nielsen}. It is worthy to mention that this quantity can be non zero for classically correlated states.    

\section{Discussions}
Quantumness in correlation is a topic of fundamental interest. It captures more general correlation than entanglement. As discussed earlier for bipartite scenario \emph{discord} is the  quantity which measures the quantumness present in a correlation. The concepts of quantum correlations can easily be extended to multi-particle scenario in the sense that a multi-partite quantum state contains quantum correlation if it can not be written as convex combination of any orthonormal product basis of the subsystems pertaining the whole system. In this work we show that this non-classical feature of correlation has manifestation in thermodynamics, particularly in work extraction from isolated systems. In this direction we have proved that presence of quantum correlations in a multi-partite state sufficiently imply non-zero difference between global and local ergotropy, which we call ergotropic gap. To motivate local ergotropy we have considered a situation where the spatially separated parties are unable to implement any global interaction field. This leads to the concept of extracting optimal work by transforming the reduced states to corresponding passive states, applying local unitaries, independently. As a future research one can try to solve the following questions. Firstly, it is interesting to classify the states for which ergotropic gap is nonzero. Categorizing the concept of ergotropy in situations where different parties are allowed to come together and can apply global unitaries also quite interesting. 

{\bf Acknowledgment}: We would like to acknowledge discussions with Prof. G. Kar. Discussion with Manabendra Nath Bera is gratefully acknowledged. We thank Tamal Guha for simulating many discussions. AM acknowledges support from the
CSIR project 09/093(0148)/2012-EMR-I.


\begin{thebibliography}{99}
\bibitem{Landauer'1961} R. Landauer, \emph{Irreversibility and Heat Generation in the Computing Process},
\href{http://ieeexplore.ieee.org/xpl/articleDetails.jsp?arnumber=5392446}{IBM J. Res. Dev. {\bf 5}, 183 (1961)}.

\bibitem{Bennett'1982} C. H. Bennett, \emph{The Thermodynamics of Computation- a Review},
\href{http://www.pitt.edu/~jdnorton/lectures/Rotman_Summer_School_2013/thermo_computing_docs/Bennett_1982.pdf}{Int. J. Theor. Phys. {\bf 21}, 905 (1982)}.

\bibitem{Vedral'2009} K. Maruyama, F. Nori, and V. Vedral, \emph{Colloquium: The physics of Maxwell’s demon and information}, \href{http://dx.doi.org/10.1103/RevModPhys.81.1}{Rev. Mod. Phys. {\bf 81}, 1-23 (2009)}.

\bibitem{Lubkin'1987} E. Lubkin, \emph{Keeping the entropy of measurement: Szilard revisited},
\href{http://link.springer.com/article/10.1007\%2FBF00670091}{Int. J. Theor. Phys. {\bf 26}, 523 (1987)}.

\bibitem{Vedral'2011} L. d. Rio,	J. Åberg,	R. Renner,	O. Dahlsten and V. Vedral, \emph{The thermodynamic meaning of negative entropy}, \href{http://www.nature.com/nature/journal/v474/n7349/full/nature10123.html}{Nature {\bf 474}, 61–63 (2011)}.

\bibitem{Spekkens'2013}F G. S. L. Brandão, M. Horodecki, J. Oppenheim, J. M. Renes, and R. W. Spekkens, \emph{Resource Theory of Quantum States Out of Thermal Equilibrium}, \href{http://dx.doi.org/10.1103/PhysRevLett.111.250404}{Phys. Rev.Lett. {\bf 111}, 250404 (2013)}.

\bibitem{Oppenheim'2013} M. Horodecki and J. Oppenheim, \emph{(Quantumness in the context of) resource theories}, \href{http://www.worldscientific.com/doi/abs/10.1142/S0217979213450197}{Int. J. Mod. Phys. B {\bf 27}, 1345019 (2013)}.
	
\bibitem{Wehner'2014} F. Brandão, M. Horodecki, N. Ng, J. Oppenheim and S.Wehner, \emph{The second laws of quantum thermodynamics}, 
\href{10.1073/pnas.1411728112}{PNAS {\bf 112}, 3275 (2015)}.

\bibitem{Gour'2015} G. Gour, M. P. M\"{u}ller, V. Narasimhachar, R. W. Spekkens, N. Y. Halpern, \emph{The resource theory of informational nonequilibrium in thermodynamics},
\href{http://www.sciencedirect.com/science/article/pii/S037015731500229X}{Phys. Rep. {\bf 583}, 1 (2015)}.

\bibitem{Fannes'2013} R. Alicki and M. Fannes, \emph{Entanglement boost for extractable work from ensembles of quantum batteries}, \href{http://dx.doi.org/10.1103/PhysRevE.87.042123}{Phys. Rev. E , {\bf 87} 042123 (2013)}.

\bibitem{Acin'2013} K. V. Hovhannisyan, M. Perarnau-Llobet, M. Huber, and A. Ac\'{c}n, \emph{Entanglement Generation is Not Necessary for Optimal Work Extraction}, 
\href{http://dx.doi.org/10.1103/PhysRevLett.111.240401}{Phys. Rev. Lett. {\bf 111}, 240401 (2013)}.

\bibitem{Skrzypczyk'2014} N. Brunner, M. Huber, N. Linden, S. Popescu, R. Silva, and P. Skrzypczyk, \emph{Entanglement enhances cooling in microscopic quantum refrigerators}, 
\href{http://dx.doi.org/10.1103/PhysRevE.89.032115}{Phys. Rev. E {\bf 89}, 032115 (2014)}.

\bibitem{Horodecki'2009} R. Horodecki, P. Horodecki, M. Horodecki, and K. Horodecki, \emph{Quantum entanglement}, \href{http://dx.doi.org/10.1103/RevModPhys.81.865}{Rev. Mod. Phys. {\bf 81}, 865 (2009)} and references therein.


\bibitem{Madhok'2011} V. Madhok and A. Datta, \emph{Interpreting quantum discord through quantum state merging},
\href{http://journals.aps.org/pra/abstract/10.1103/PhysRevA.83.032323}{Phys. Rev. A {\bf 83}, 032323 (2011)}.

\bibitem{Cavalcanti'2011} D. Cavalcanti, L. Aolita, S. Boixo, K. Modi, M. Piani, and A. Winter, \emph{Operational interpretations of quantum discord},
\href{http://journals.aps.org/pra/abstract/10.1103/PhysRevA.83.032324}{Phys. Rev. A {\bf 83}, 032324 (2011)}.

\bibitem{Dakic'2012} B. Daki\'{c} \emph{et al}, \emph{Quantum discord as resource for remote state preparation}, \href{http://www.nature.com/nphys/journal/v8/n9/full/nphys2377.html}{Nat. Phys.	{\bf 8}, 666 (2012)}.

%\bibitem{Zurek'2001} H. Ollivier and W. H. Zurek, \emph{Quantum Discord: A Measure of the Quantumness of Correlations}, \href{http://dx.doi.org/10.1103/PhysRevLett.88.017901}{Phys. Rev. Lett. {\bf 88}, 017901 (2001)}.

%\bibitem{Henderson'2001} L. Henderson and V. Vedral, \emph{Classical, quantum and total correlations},
%\href{http://iopscience.iop.org/0305-4470/34/35/315/}{J. Phys. A: Math. Gen. {\bf 34}, 6899 (2001)}.

\bibitem{Vedral'2012} K. Modi, A. Brodutch, H. Cable, T. Paterek, and V. Vedral, \emph{The classical-quantum boundary for correlations: Discord and related measures}, 
\href{http://dx.doi.org/10.1103/RevModPhys.84.1655}{Rev. Mod. Phys. {\bf 84}, 1655 (2012)} and references therein.

\bibitem{Lifshitz'1978} L.D. Landau and E. M. Lifshitz, Statistical Physics, I (Pergamon Press, Oxford) 1978.

\bibitem{Callen'1985} H. B. Callen, Thermodynamics (John Wiley, New York) 1985.

\bibitem{Woronowicz'1978} W. Pusz and S. L. Woronowicz, \emph{Passive states and KMS states for general quantum systems}, \href{https://projecteuclid.org/euclid.cmp/1103901491}{Commun. Math. Phys. {\bf 58}, 273 (1978)}.

\bibitem{Lenard'1978} A. Lenard, \emph{Thermodynamical proof of the Gibbs formula for elementary quantum systems}, 
\href{http://link.springer.com/article/10.1007/BF01011769}{J. Stat. Phys. {\bf 19}, 575 (1978)}.
	
\bibitem{Nieuwenhuizen'2004} A. E. Allahverdyan , R. Balian and T. M. Nieuwenhuizen,\emph{Maximal work extraction from finite quantum systems}, 
\href{http://iopscience.iop.org/0295-5075/67/4/565}{Europhys. Lett., {\bf 67} (4), 565(2004)}

\bibitem{Aberg'2013} J. \AA{}berg, \emph{Truly work-like work extraction via a single-shot analysis},
\href{http://www.nature.com/ncomms/2013/130626/ncomms2712/full/ncomms2712.html}{Nature Communications {\bf4}, 1925 (2013)}.

\bibitem{Horodecki'2013} M. Horodecki and J. Oppenheim, \emph{Fundamental limitations for quantum and nanoscale thermodynamics}, 
\href{http://www.nature.com/ncomms/2013/130626/ncomms3059/full/ncomms3059.html}{Nature Communications 4, 2059 (2013)}.

\bibitem{Popescu'2014} P. Skrzypczyk, A. J. Short and S. Popescu, \emph{Work extraction and thermodynamics for individual quantum systems}, 
\href{http://www.nature.com/ncomms/2014/140627/ncomms5185/abs/ncomms5185.html}{Nat. Comm. {\bf 5}, 4185 (2014)}.

\bibitem{Aberg'2014} J. \AA{}berg, \emph{Catalytic Coherence},
\href{http://journals.aps.org/prl/abstract/10.1103/PhysRevLett.113.150402}{Phys. Rev. Lett. {\bf 113}, 150402 (2014)}.

\bibitem{Lostaglio'2015} M. Lostaglio, K. Korzekwa, D. Jennings, and T. Rudolph, \emph{Quantum Coherence, Time-Translation Symmetry, and Thermodynamics},
\href{http://journals.aps.org/prx/abstract/10.1103/PhysRevX.5.021001}{Phys. Rev. X {\bf5}, 021001 (2015)}.

\bibitem{Acin'2014} M. Perarnau-Llobet, K. V. Hovhannisyan, M. Huber, P. Skrzypczyk, N. Brunner, and A. Ac\'{\i}n, \emph{Extractable work from correlations}, 
\href{http://journals.aps.org/prx/abstract/10.1103/PhysRevX.5.041011}{Phys. Rev. X {\bf 5}, 041011 (2015)}.

\bibitem{Brunner'2015} P. Skrzypczyk, R. Silva, and N. Brunner, \emph{Passivity, complete passivity, and virtual temperatures},
\href{http://dx.doi.org/10.1103/PhysRevE.91.052133}{Phys. Rev. E {\bf 91}, 052133 (2015)}.

\bibitem{Brunner'2014} N. Brunner, D. Cavalcanti, S. Pironio, V. Scarani, and S. Wehner, \emph{Bell nonlocality},
\href{http://journals.aps.org/rmp/abstract/10.1103/RevModPhys.86.419}{Rev. Mod. Phys. {\bf 86}, 839 (2014)} and references therein.

\bibitem{Wiseman'2007} H. M. Wiseman, S. J. Jones, and A. C. Doherty, \emph{Steering, Entanglement, Nonlocality, and the Einstein-Podolsky-Rosen Paradox},
\href{http://journals.aps.org/prl/abstract/10.1103/PhysRevLett.98.140402}{Phys. Rev. Lett. {\bf 98}, 140402 (2007)}.

\bibitem{application1} A. K. Ekert, \emph{Quantum cryptography based on Bell’s theorem},
\href{http://journals.aps.org/prl/abstract/10.1103/PhysRevLett.67.661}{Phys. Rev. Lett. {\bf 67}, 661 (1991)}.

\bibitem{application2} C. H. Bennett, G. Brassard, and N. D. Mermin, \emph{Quantum cryptography without Bell’s theorem}
\href{http://journals.aps.org/prl/abstract/10.1103/PhysRevLett.68.557}{Phys. Rev. Lett. {\bf 68}, 557 (1992)}.

\bibitem{application3} C. H. Bennett and S. J. Wiesner, \emph{Communication via one- and two-particle operators on Einstein-Podolsky-Rosen states}
\href{http://journals.aps.org/prl/abstract/10.1103/PhysRevLett.69.2881}{Phys. Rev. Lett. {\bf 69}, 2881 (1992)}.

\bibitem{application4} C. H. Bennett, G. Brassard, C. Crepeau, R. Jozsa, A. Peres, and W. K. Wootters, \emph{Teleporting an unknown quantum state via dual classical and Einstein-Podolsky-Rosen channels}
\href{http://journals.aps.org/prl/abstract/10.1103/PhysRevLett.70.1895}{Phys. Rev. Lett. {\bf 70}, 1895 (1993)}.

\bibitem{application5} J. Barrett, L. Hardy, and A. Kent, \emph{No Signaling and Quantum Key Distribution},
\href{http://journals.aps.org/prl/abstract/10.1103/PhysRevLett.95.010503}{Phys. Rev. Lett. {\bf 95}, 010503 (2005)}.

\bibitem{application6} N. Brunner, S. Pironio, A. Acin, N. Gisin, A. Allan Methot, and V. Scarani, \emph{Testing the Dimension of Hilbert Spaces},
\href{http://journals.aps.org/prl/abstract/10.1103/PhysRevLett.100.210503}{Phys. Rev. Lett. {\bf 100}, 210503 (2008)}.

\bibitem{application7} S. Pironio \emph{et al.}, \emph{Random numbers certified by Bell’s theorem},
\href{http://www.nature.com/nature/journal/v464/n7291/full/nature09008.html}{Nature {\bf 464}, 1021 (2010)}.

\bibitem{application8} S. Das, M. Banik, A. Rai, MD R. Gazi, and S. Kunkri, \emph{Hardy's nonlocality argument as a witness for postquantum correlations},
\href{http://journals.aps.org/pra/abstract/10.1103/PhysRevA.87.012112}{Phys. Rev. A {\bf 87}, 012112 (2013)}.

\bibitem{application9} A. Mukherjee, A. Roy, S. S. Bhattacharya, S. Das, Md. R. Gazi, and M. Banik, \emph{Hardy's test as a device-independent dimension witness},
\href{https://journals.aps.org/pra/abstract/10.1103/PhysRevA.92.022302}{Phys. Rev. A 92, 022302 (2015)}.

\bibitem{application10} M. Banik, S. S. Bhattacharya, A. Mukherjee, A. Roy, A. Ambainis, and A. Rai, \emph{Limited preparation contextuality in quantum theory and its relation to the Cirel'son bound},
\href{https://journals.aps.org/pra/abstract/10.1103/PhysRevA.92.030103}{Phys. Rev. A {\bf 92}, 030103(R) (2015)}.

\bibitem{application11} A. Chaturvedi  and M. Banik, \emph{Measurement-device-independent randomness from local entangled states}, 
\href{http://iopscience.iop.org/article/10.1209/0295-5075/112/30003/meta;jsessionid=E5B18E7672D6C5C61FBE37DBD2513C25.c2.iopscience.cld.iop.org}{EPL, {\bf 112}, 30003 (2015)}.

\bibitem{application12} A. Roy, A. Mukherjee, T. Guha, S. Ghosh, S. S. Bhattacharya, M. Banik, \emph{Nonlocal correlations: Fair and Unfair Strategies in Bayesian Game},
\href{http://arxiv.org/abs/1601.02349}{arXiv:1601.02349v2}. 

%\bibitem{Madhok'2011} V. Madhok and A. Datta, \emph{Interpreting quantum discord through quantum state merging},
%\href{http://journals.aps.org/pra/abstract/10.1103/PhysRevA.83.032323}{Phys. Rev. A {\bf 83}, 032323 (2011)}.

%\bibitem{Cavalcanti'2011} D. Cavalcanti, L. Aolita, S. Boixo, K. Modi, M. Piani, and A. Winter, \emph{Operational interpretations of quantum discord},
%\href{http://journals.aps.org/pra/abstract/10.1103/PhysRevA.83.032324}{Phys. Rev. A {\bf 83}, 032324 (2011)}.

%\bibitem{Dakic'2012} B. Daki\'{c} \emph{et al}, \emph{Quantum discord as resource for remote state preparation}, %\href{http://www.nature.com/nphys/journal/v8/n9/full/nphys2377.html}{Nat. Phys.	{\bf 8}, 666 (2012)}.

%\bibitem{Appendix} See the appendix.

\bibitem{Groisman'2007} B. Groisman, D. Kenigsberg, T. Mor, \emph{``Quantumness" versus ``Classicality" of Quantum States},
\href{http://arxiv.org/abs/quant-ph/0703103}{arXiv:quant-ph/0703103}.

\bibitem{Horodecki'1995} R. Horodecki, P. Horodecki, M. Horodecki, \emph{Violating Bell inequality by mixed spin-$\frac{1}{2}$ states: necessary and sufficient condition},
\href{http://www.sciencedirect.com/science/article/pii/037596019500214N}{Phys. Lett. A {\bf 200}, 340 (1995)}.

\bibitem{nielsen} M. A. Nielsen and I. L. Chuang \emph{Quantum Computation and Quantum Information} (New York:
Cambridge University Press, 2011).
\end{thebibliography}
\end{document}